\RequirePackage{amsthm}

\documentclass[sn-mathphys,Numbered]{sn-jnl}

\usepackage{graphicx}%
\usepackage{multirow}%
\usepackage{amsmath,amssymb,amsfonts}%
\usepackage{amsthm}%
\usepackage{mathrsfs}%
\usepackage[title]{appendix}%
\usepackage{xcolor}%
\usepackage{textcomp}%
\usepackage{manyfoot}%
\usepackage{booktabs}%
\usepackage{algorithm}%
\usepackage{algorithmicx}%
\usepackage{algpseudocode}%
\usepackage{listings}%

\usepackage{subcaption}

\theoremstyle{thmstyleone}%
\theoremstyle{thmstyletwo}%

\theoremstyle{thmstylethree}%

\begin{document}

\title[Full-Scale Readout Electronics for the ECHo-100k Experiment]{Full-Scale Readout Electronics for the ECHo Experiment}


\author*[1]{\fnm{Timo} \sur{Muscheid}}\email{timo.muscheid@kit.edu}

\author[1]{\fnm{Robert} \sur{Gartmann}} 
\author[1]{\fnm{Nick} \sur{Karcher}}
\author[2]{\fnm{Felix} \sur{Schuderer}}
\author[2]{\fnm{Martin} \sur{Neidig}}
\author[1]{\fnm{Luis E.} \sur{Ardila-Perez}} 
\author[1]{\fnm{Matthias} \sur{Balzer}}
\author[2,1]{\fnm{Sebastian} \sur{Kempf}}
\author[1]{\fnm{Oliver} \sur{Sander}}

\affil[1]{\orgdiv{Institute for Data Processing and Electronics (IPE)}, \orgname{Karlsruhe Institute of Technology (KIT)}, \orgaddress{\country{Germany}}}

\affil[2]{\orgdiv{Institute of Micro- and Nanoelectronic Systems (IMS)}, \orgname{Karlsruhe Institute of Technology (KIT)}, \orgaddress{\country{Germany}}}

\abstract{Recent advances in the development of cryogenic particle detectors such as magnetic microcalorimeters (MMCs) allow the fabrication of sensor arrays with an increasing number of pixels. Since these detectors must be operated at the lowest temperatures, the readout of large detector arrays is still quite challenging. This is especially true for the ECHo experiment, which presently aims to simultaneously run 6,000 two-pixel detectors to investigate the electron neutrino mass. For this reason, we developed a readout system based on a microwave SQUID multiplexer ($\mu$MUX) that is operated by a custom software-defined radio (SDR) at room-temperature.
The SDR readout electronics consist of three distinct hardware units: a data processing board with a Xilinx ZynqUS+ MPSoC; a converter board that features DACs, ADCs, and a coherent clock distribution network; and a radio frequency front-end board to translate the signals between the baseband and the microwave domains. Here, we describe the characteristics of the full-scale SDR system. First, the generated frequency comb for driving the $\mu$MUX was evaluated. Subsequently, by operating the SDR in direct loopback, the crosstalk of the individual channels after frequency demultiplexing was investigated. Finally, the system was used with a 16-channel $\mu$MUX to evaluate the linearity of the SDR, and the noise contributed to the overall readout setup.}




\maketitle

\section{Introduction}\label{sec:introduction}

Experiments in various fields of physics, possibly containing hundreds of thousands of sensors, require state-of-the-art readout electronics capable of handling the ever-increasing data rates and performing real-time data analysis. Detectors working at cryogenic temperatures further complicate the readout architecture since the number of cables interfacing the cryogenic domain and the power dissipation at cryogenic temperatures need to be minimized. Both challenges can be tackled by multiplexing techniques such as time division multiplexing (TDM)\,\cite{Doriese2015} or frequency division multiplexing (FDM)\,\cite{IRWIN2006802}. Next-generation projects are evaluating new opportunities made possible by recent developments in this field. Readout via frequency multiplexed magnetic microcalorimeters (MMC) is in particular assessed within the ECHo experiment \cite{Gastaldo2017}. ECHo aims to investigate the mass of the electron neutrino by measuring the electron capture process of Holmium-163\,\cite{Gastaldo2017}. The released energy is measured by MMCs connected to a microwave SQUID multiplexer ($\mu$MUX)\,\cite{Wegner2018}. Each $\mu$MUX consists of 400 resonators with resonance frequencies between 4 and 8\,GHz and two MMCs each. In total, 800 detectors will be multiplexed on a single transmission line. 
In recent years, the readout concept, including the room-temperature electronics for online data analysis, has been demonstrated with various prototype iterations\,\cite{Sander2019, Gartmann2022}. Now, the system has been scaled up to target the 12,000 detectors in the experiment. 
In this work, the full-scale room-temperature software-defined radio (SDR) system is presented and characterized.

\section{The Full-Scale SDR System}\label{sec:sdr}

\begin{figure}[h]
\centering
\begin{subfigure}{.3\textwidth}
  \centering
  \includegraphics[height=2.9cm]{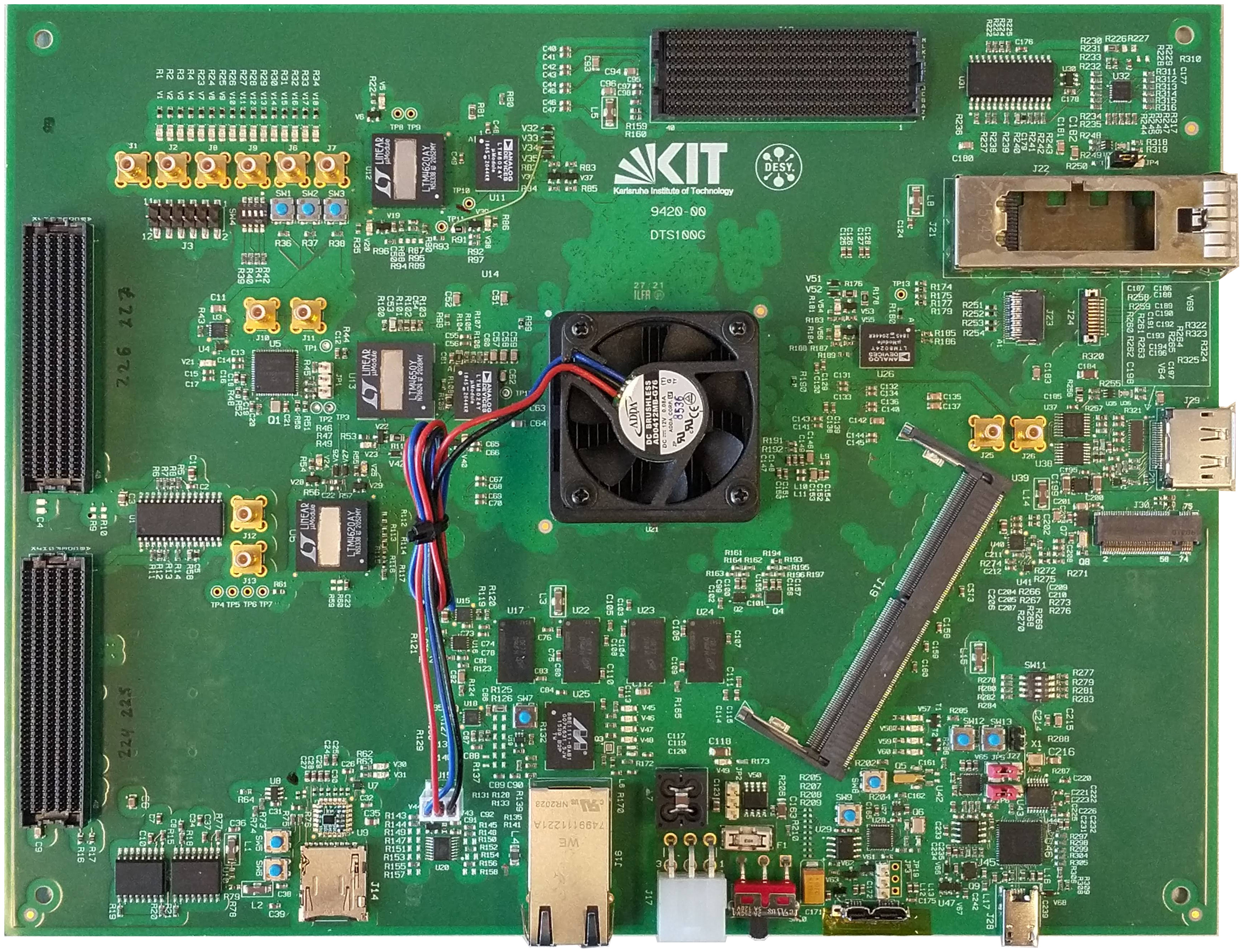}
\end{subfigure}
\hfill
\begin{subfigure}{.34\textwidth}
  \centering
  \includegraphics[height=2.9cm]{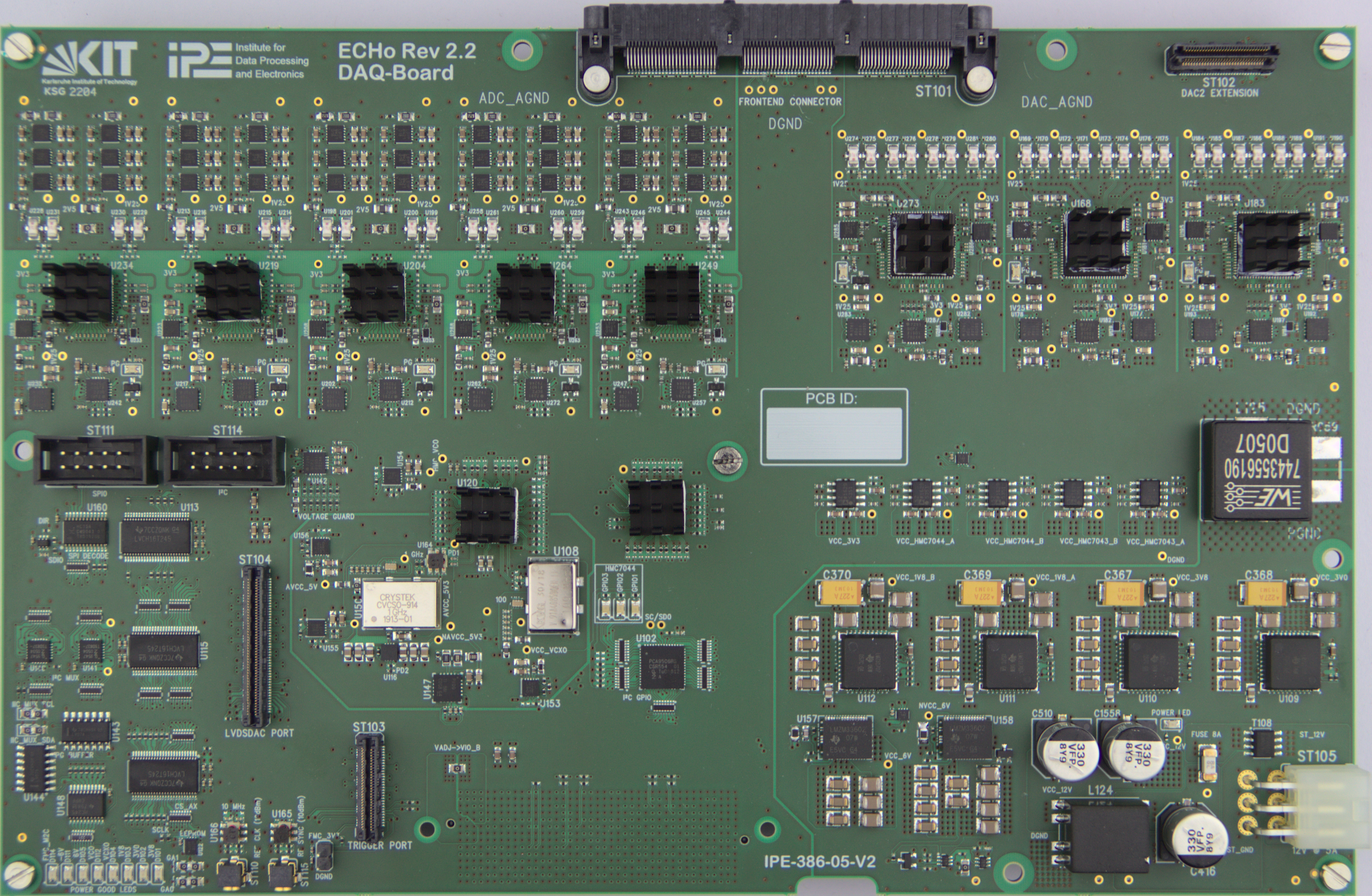}
\end{subfigure}
\hfill
\begin{subfigure}{.34\textwidth}
  \centering
  \includegraphics[height=2.9cm]{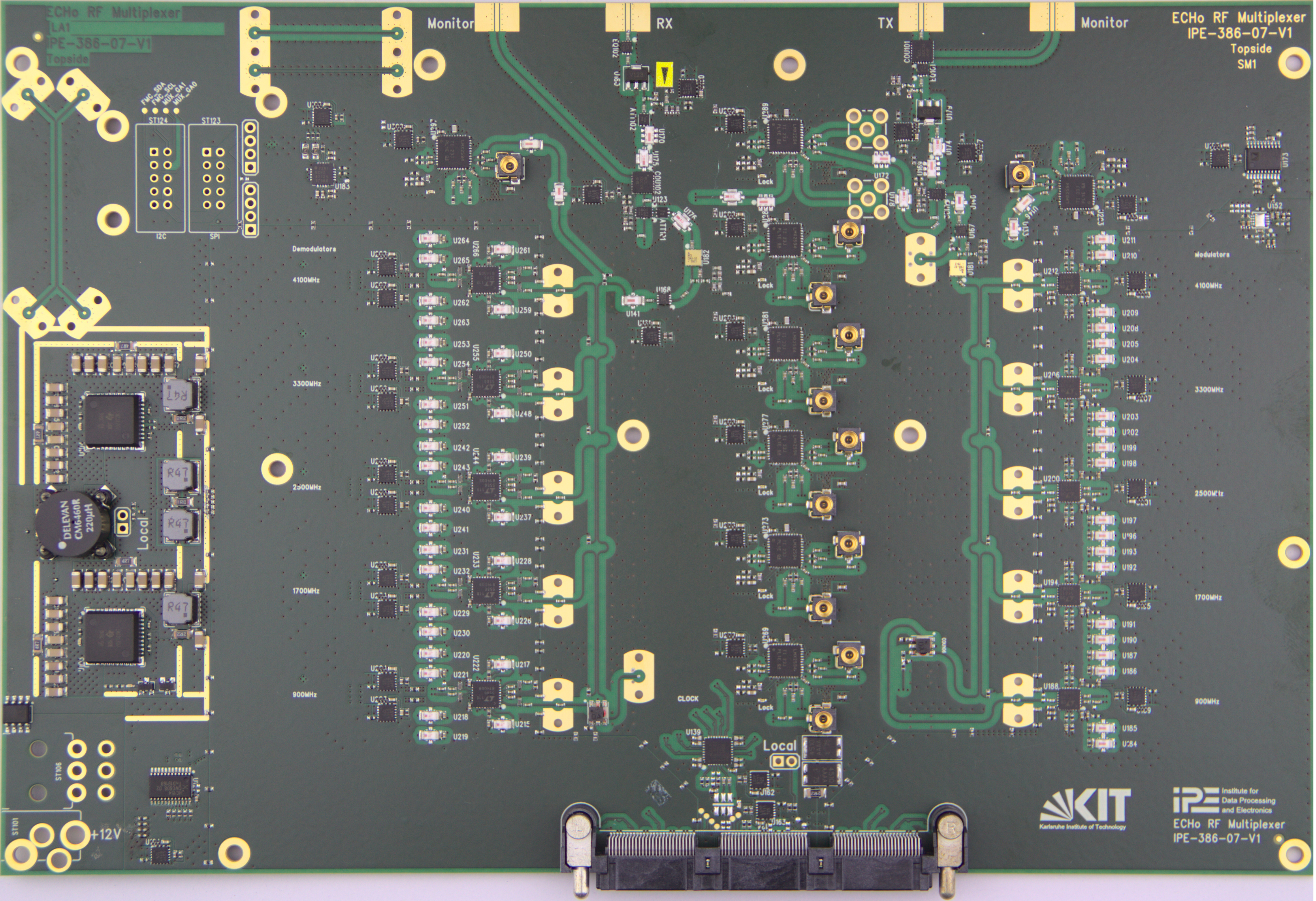}
\end{subfigure}
\caption{Full-stack ECHo DAQ SDR electronics: digital processing board (left); analog/digital conversion board (center); high-frequency analog heterodyne mixer board (right).}
\label{fig:electronics}
\end{figure}

The room-temperature readout system\,\cite{Sander2019} follows a modular approach and consists of three distinct hardware units as seen in \autoref{fig:electronics}: a custom MPSoC board for digital signal processing\,\cite{Muscheid_2023}, a conversion stage, and a heterodyne mixing stage\,\cite{Gartmann2022}. 

\begin{figure}[h]
  \centering
  \includegraphics[width=\linewidth]{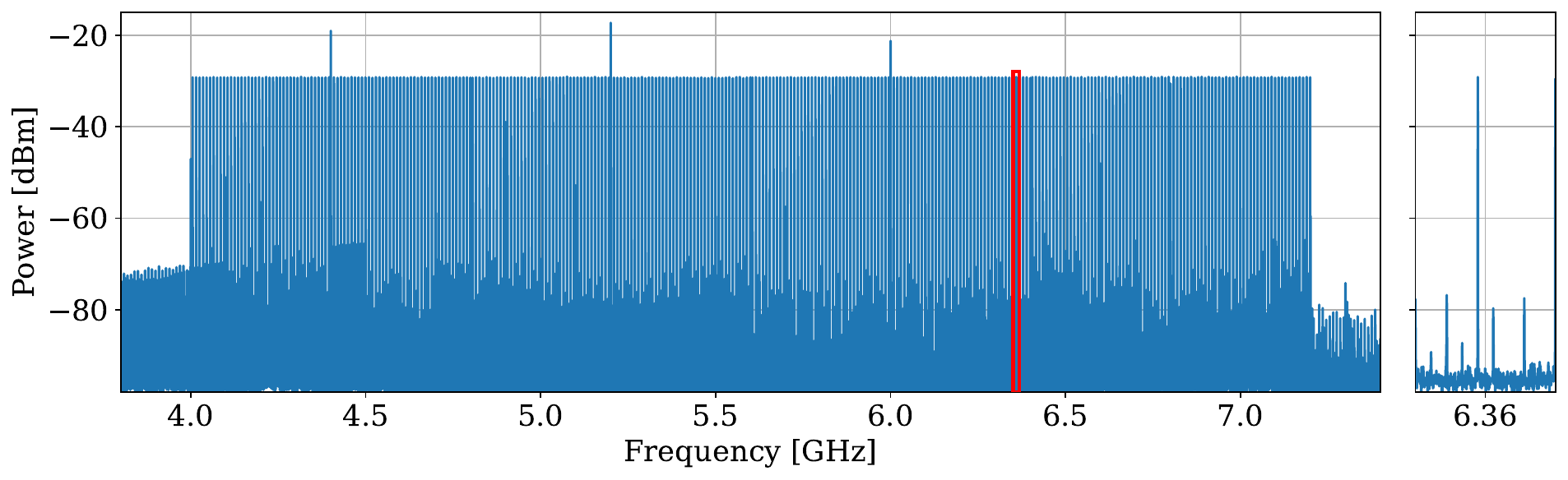}
  \caption{Frequency comb containing 4 of 5 subbands with 80 tones each, where the IQ imbalance is corrected and the power is equalized for all tones. A selected tone is highlighted and zoomed in.
  }
  \label{fig:comb}
\end{figure}

The target frequency band between 4 and 8\,GHz is divided into five subbands of 800\,MHz each that operate independently from each other. The I and Q components of the frequency comb for resonator stimulation are generated digitally on the Programmable Logic (PL) and converted to the analog domain by DACs operated at 1\,GHz. These baseband signals are subsequently shifted to individual bandbass signals by IQ mixers. When combined, they form a unified frequency comb, as seen in \autoref{fig:comb}. On the receiving side, the modulated frequency comb is divided and downmixed similarly into five bands, sampled separately at 1\,GHz by ADCs. The firmware running on the PL processes the five parallel data streams and performs channelization, flux-ramp demodulation, and event detection\,\cite{Karcher2022}. The resulting data streams containing the detector samples can be stored on the platform itself, or they can be sent to a client PC for post-processing and spectrum analysis.

\section{System characterization}\label{sec:characterization}

Following final commissioning, the full-scale room-temperature readout electronics system was characterized. The measurement setups and results are presented in the following sections.

\subsection{Frequency comb generation}
The frequency comb generated by the SDR contains all the resonance frequencies needed to drive the multiplexer. The image rejection ratio (IRR) of the single tones is of main interest. Strong image frequencies increase the total power sent to the cryostat, potentially bring the HEMT amplifier into saturation, and are a severe source of crosstalk. These images are generated by phase and amplitude deviations of the I and Q components resulting from mismatches in the signal trace lengths from the DAC to the IQ mixer and component variation. By intentionally generating I and Q data with slightly different phases and amplitudes, these errors can be compensated.
The TX-output of the SDR electronics was connected to a spectrum analyzer (Rohde \& Schwarz FSWP50), and a frequency comb containing 80 random tones per subband was generated. After an IQ imbalance correction\,\cite{ad_2009}, the system reached an IRR $>40$\,dB; then limited by intermodulation and distortion in the following analog signal path. To obtain a flat comb, the weakest tone across the complete frequency band was detected, and the amplitude of all other tones was reduced to match that power. The signal power of individual tones can be adjusted by the users to match the requirements of the experiment. For ECHo in particular, the $\mu$MUX, is designed for a signal power of -70\,dBm per resonator\,\cite{Richter2021}. The signal conditioning in the cryostat requires an input power of -27\,dBm per tone at the cryogenic interface for optimal performance\,\cite{Ahrens2022}. This requirement is being fulfilled by the presented system.

\subsection{Channel separation}
\begin{figure}[h]
\centering%
\begin{subfigure}{.48\textwidth}
  \centering%
  \includegraphics[height=3.25cm]{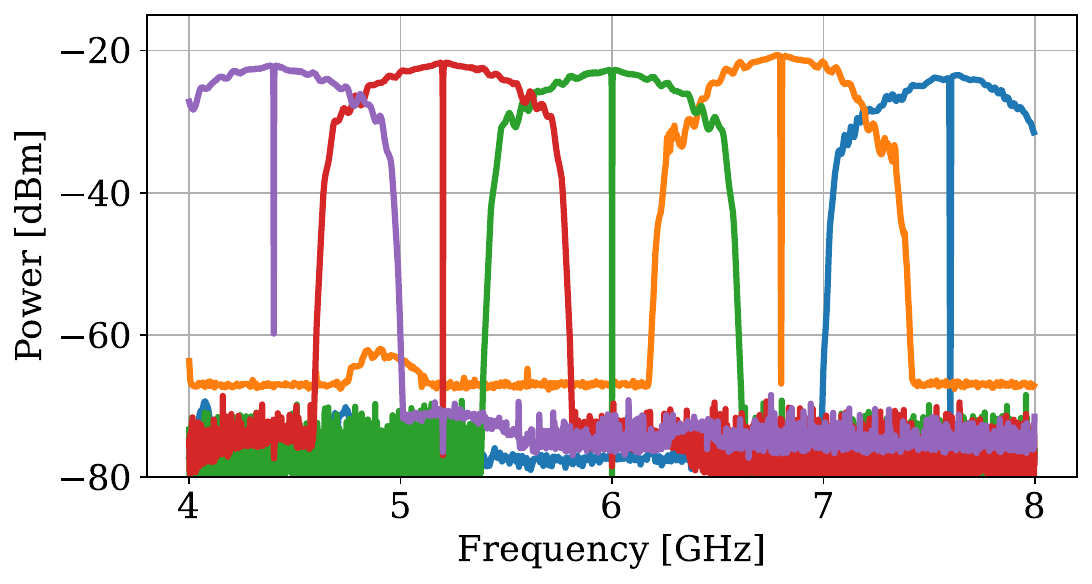}%
  \label{fig:analog_crosstalk}
\end{subfigure}%
\hfill
\begin{subfigure}{.48\textwidth}
  \centering%
  \includegraphics[height=3.25cm]{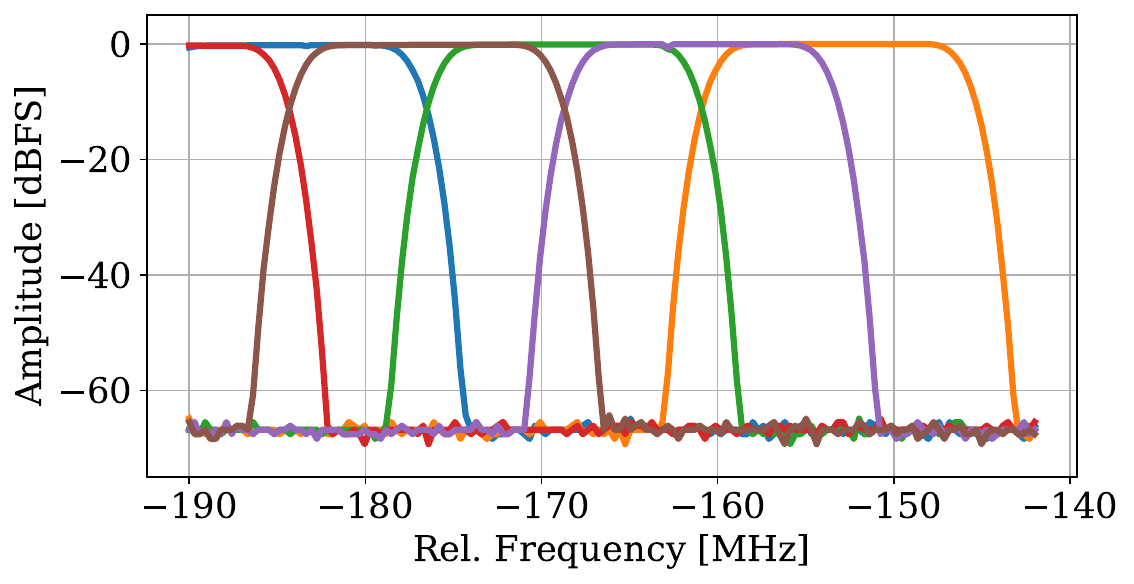}%
  \label{fig:digital_crosstalk}
\end{subfigure}
\caption{Evaluation of the channel separation. First, the analog wideband comb is separated into 5 subbands with 800\,MHz each (left). Then, a polyphase filter bank within the digital domain channelizes individual tones into a TDM scheme. For simplicity, only an extract of the digital filter bank for one data stream is shown (right).}
\label{fig:crosstalk}
\end{figure}

As described in \autoref{sec:sdr}, the wideband frequency comb is first divided into five subbands of 800\,MHz each and mixed to the baseband. 
This analog band separation is evaluated by sweeping a single tone from 4 to 8\,GHz and measuring the five baseband frequency responses of the standalone RF frontend. By applying an FFT, the strongest signal component for each input frequency was identified. The result of this measurement is shown on the left side of \autoref{fig:crosstalk}. Feedthrough of the clock reference was not sufficiently filtered by demodulator 4, leading to a higher noise floor in that subband. Similarly, the harmonic mixing around 5\,GHz can be attenuated further by adding another lowpass filter. Demodulator 5 originally also processed signals outside of its responsible frequency band. However, this issue was solved by introducing a low-pass filter before its intermediate frequency input. A minor revision of the RF frontend board will include the necessary modifications for optimization.

The ADCs provide internal digital down-conversion (DDC) and convert each analog baseband signal into two separate data streams with sampling rates of 500\,MHz each, containing the upper and lower sidebands, respectively. 
The digital channelization is performed by a polyphase channelizer module that separates the 500\,MHz signal into 32 equidistant channels with a passband of 11\,MHz. Additionally, a second parallel polyphase channelizer module converts the frequency-shifted copy of the input signal into 32 channels that have their center frequency in the stop-band of the first channelizer. In total, the digital channelization converts the 10 parallel 500\,MHz bands into 64 overlapping channels each. Subsequently, a tunable numerically controlled oscillator (NCO) and a lowpass filter with a cutoff frequency of 1.6\,MHz are used on each channel to mix the carrier frequencies to DC. This specific cutoff frequency was chosen as a compromise between detector bandwidth and resource consumption, which increases with higher orders of the filter. The digital channel separation has been evaluated by generating a single tone in a room-temperature loopback configuration and measuring the output amplitudes of every channel. Across the full frequency band, digital crosstalk attenuation of more than 55\,dB is achieved. An extract of the crosstalk after this channelization is shown on the right side of \autoref{fig:crosstalk}.

\subsection{Noise contribution}\label{subsec:noise}
After demodulation of the carrier tones, the flux-ramp modulated SQUID signal in form of a low-frequency sine-like signal remains. Applying flux-ramp demodulation\,\cite{Karcher2022} to this signal allows the reconstruction of the raw detector signal. However, without a detector as phase-modulating component, the flux-ramp demodulated signal shows the in-system noise. The noise power spectral density (PSD) was compared in two scenarios: a direct room-temperature loopback and a full system with connection to the cryogenic domain, including a preliminary stack-up of cryogenic and room-temperature amplifiers for boosting the signal power prior to the processing stage. The cryogenic setup contained the $\mu$MUX with SQUIDs that were flux-ramp modulated. A single tone was generated to drive one resonator, and samples from the appropriate channel were acquired. In the room-temperature loopback setup, the carrier tone has been digitally amplitude-modulated in the tone generation module to mimic the flux-ramp modulated SQUID response. The left side of \autoref{fig:cryo_characterization} shows that the noise level of the room-temperature loopback is below the total noise of the cryogenic setup. Therefore, the room-temperature electronics are not the limiting factor for the measurement sensitivity. Due to the 1.6\,MHz lowpass filter within the channelization stage, the power of the noise above the cutoff frequency is strongly attenuated.

\subsection{System linearity}\label{subsec:linearity}
For verification of the readout electronics under real conditions, it is required to include the $\mu$MUX with sensing SQUIDs and phase-modulating components in the signal path. To obtain a controlled measurement environment, we connected the input coil of each $\mu$MUX SQUID to an external signal line via a superconducting flux transformer. This galvanic isolation was required since a direct connection strongly affected the $\mu$MUX resonator quality. This MMC emulator setup allows us to inject known signals into individual $\mu$MUX channels, e.g. to investigate the overall system linearity.

\begin{figure}[h]
\centering
\begin{subfigure}{.48\textwidth}
  \centering%
  \includegraphics[width=\linewidth]{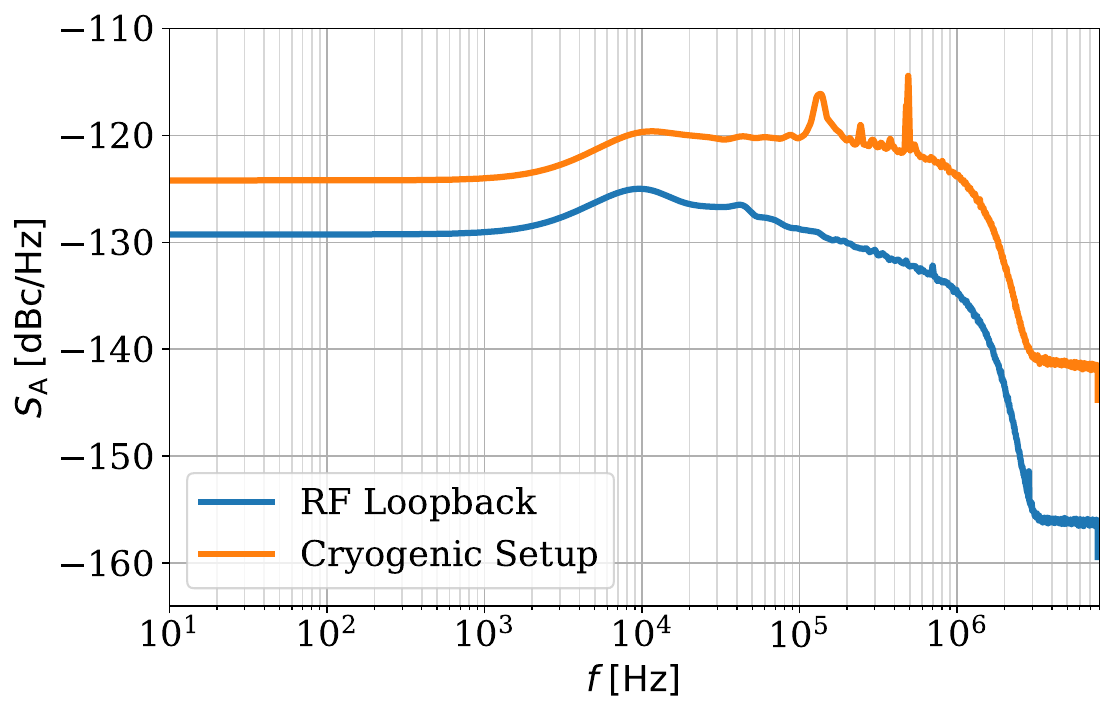}%
  \label{fig:noise}
\end{subfigure}%
\hfill
\begin{subfigure}{.48\textwidth}
  \centering%
  \includegraphics[width=\linewidth]{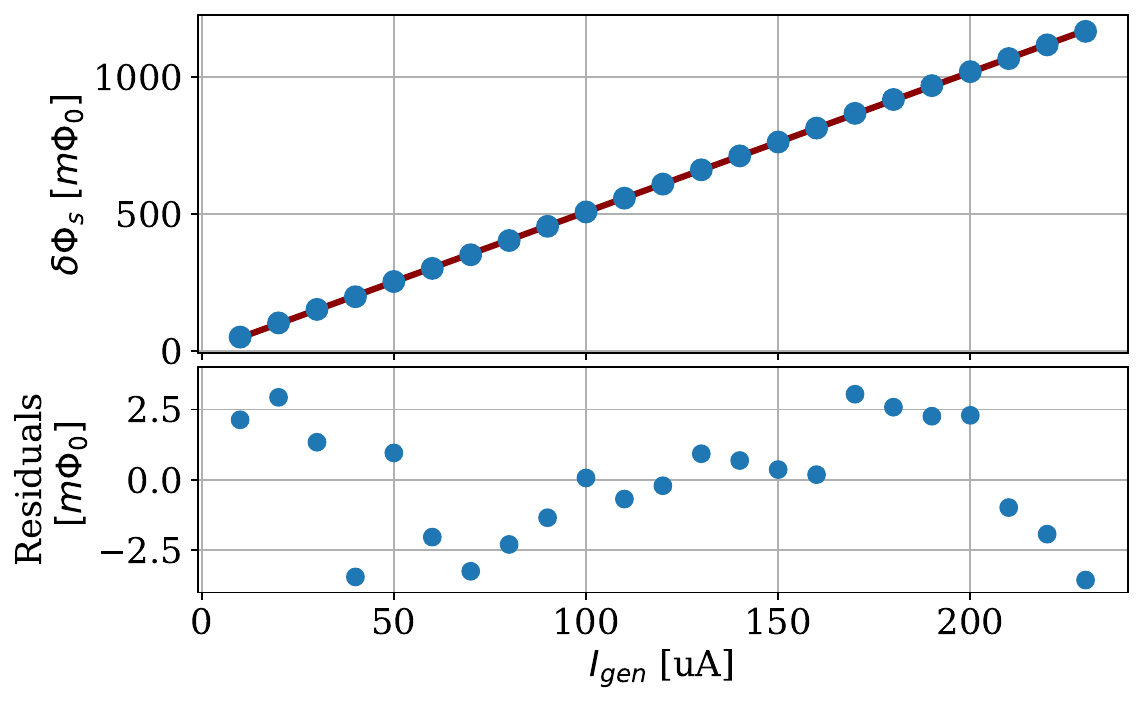}%
  \label{fig:linearity}
\end{subfigure}
\caption{Full SDR system characterization connected to the cryogenic domain. Left: power spectral density (PSD) of the noise amplitude on a single demodulated channel. PSD is normalized to the carrier tone power to compensate for different input powers at the ADC in the two setups due to attenuation by cryostat electronics. Since the noise level of the RF loopback is below the full cryogenic setup, the SDR electronics do not significantly influence the measurement accuracy. Right: linearity on a single channel measured with a series of triangular input signals of increasing amplitude. The SDR response was calculated to $194 \pm 0.002 \mu A/ \Phi_{0}$.}
\label{fig:cryo_characterization}
\end{figure}

By using this MMC emulator, the system linearity of the SDR was evaluated. A linear response of the electronics is a key requirement to guarantee accurate measurement of the particle energy in the experiment. The linearity of the system is determined by taking consecutive measurements with increasing signal amplitude on the MMC emulator input. The input signal is then compared to the signal amplitude calculated by the SDR after channelization and flux-ramp demodulation. For this measurement, the Magnicom SQUID electronics\,\cite{Drung_2006} were connected to the MMC emulator as a signal source. The SQUID electronics were configured to generate sinusoidal and triangular signals with amplitudes up to $260 \mu A$ to make sure that at least one magnetic flux quantum ($\Phi_{0}$) is surpassed. The SDR response has a slope of $194 \pm 0.002 \mu A/ \Phi_{0}$ as shown in \autoref{fig:cryo_characterization} on the right side. This measurement has been verified with a VNA (Keysight P5024B) as the room-temperature readout system. Post-processing for reconstruction of the MMC emulator output resulted in a slope of $187 \mu A/ \Phi_{0}$. This deviation in the slope is not fully understood yet and needs further investigation. 

\section{Conclusion}
While several prototypes of the SDR electronics had proved the readout concept successfully, the complete system with all components of signal generation and online data processing for 800 detectors has not been operated and evaluated yet. The system presented in this work allows the characterization of the interacting effects of up- and down-mixing of multiple parallel frequency bands together with analog and digital filtering for the first time ever. Analysis of the frequency comb testifies that the tones can be generated to match the multiplexer design, and the requirements of the cryogenic domain are fully met with an image rejection ratio and a signal-to-noise ratio above 40\,dB. On the RX-side, the measurement results show excellent crosstalk attenuation at the analog and digital channel separation stages below 55\,dB and an overall low noise contribution of the SDR. The measured linearity of the full system connected to the cryogenic domain at $194 \pm 0.002 \mu A/ \Phi_{0}$ promises accurate results once this system is utilized in the experiment. Overall, it was shown that readout of $\mu$MUX signals is possible with high accuracy, even for a large number of detectors.

\vfill
\backmatter
\bmhead{Acknowledgments}
The authors acknowledge the support by the Doctoral School "Karlsruhe School of Elementary and Astroparticle Physics: Science and Technology"

\bibliography{sn-bibliography}

\end{document}